\documentclass[12pt]{extarticle}
\usepackage[utf8x]{inputenc}
\usepackage{mathpazo}
\usepackage{amssymb,amsmath,amsthm}
\usepackage{fullpage}
\usepackage{hyperref}
\usepackage{graphicx}
\usepackage{calc}
\usepackage{xcolor}
\usepackage{eurosym}
\usepackage{tikz}
\usepackage{calc}
\usepackage{float}
\usepackage{caption}
\usepackage{import}
\usepackage{comment}

\everymath{\color{blue}}

\usepackage{pgfplots}
\usepackage{hyperref}
\usepackage{calligra}
\usepackage[T1]{fontenc}
\usepackage{mathrsfs}
\usepackage{centernot}
\usepackage{cancel}
\definecolor{FigBlue}{rgb}{0.150,0.60,0.80}
\definecolor{BubBlue}{rgb}{0.35,0.8,1.00}
\definecolor{DarkBlue}{rgb}{0.175,0.4,0.5}

\usepackage{enumerate}
\usepackage{pgfplots}
\usepackage{mathrsfs}
\usepackage{calligra}
\usepackage[
backend=biber,
style=authoryear,
sorting=nyt
]{biblatex}
\theoremstyle{remark}

\newtheorem*{theorem*}{Theorem}

\newtheorem*{hypothesis*}{Hypothesis}

\newtheorem{structural assumption}{Structural Assumption}
\newtheorem*{proposition*}{Proposition}
\newtheorem{proposition}{Proposition}

\newcommand\blfootnote[1]{
    \begingroup
    \renewcommand\thefootnote{}\footnote{#1}
    \addtocounter{footnote}{-1}
    \endgroup
}

\begin{document}
	\title{Counterexamples to "Transitive Regret"}
	\author{Yuan Chang\footnote{YuanChang525@outlook.com}\\
 Shuo Li Liu\footnote{shl@alumni.harvard.edu}
 }
	\date{June 10, 2024}
	\maketitle	
	\begin{abstract}
Theorem 1 in Bikhchandani \& Segal (2011) suggests that a complete, transitive, monotonic, and continuous preference is regret based if and only if it is expected utility. Their Proposition 1 suggests that transitivity and continuity of a regret-based preference implies an equivalence condition: if random variables $X$ and $Y$ have the same distribution, then $X\sim Y$. We give counterexamples to Proposition 1. \blfootnote{We are in debt to Jingni Yang for comments. All errors are ours.}
		\bigskip
	\end{abstract}
	\setcounter{page}{0}
	\thispagestyle{empty}
\setcounter{page}{1}

\section{Notations}
Let $L$ be the set of real finite-valued random variables over $(S, \Sigma, P)$ with $S = [0, 1]$, $\Sigma$ being the Borel $\sigma$-algebra, $P$ being the Lebesgue measure, and the set of outcomes being the bounded interval $[\underline{x}, \bar{x}]$. The decision maker has a preference relation $\succsim$ over $L$. Denote events by $S_i$.

\noindent\textbf{Definition 1.} The continuous function $\psi : [\underline{x}, \bar{x}] \times [\underline{x}, \bar{x}] \rightarrow \mathbb{R}$ is a \textit{regret function} if for all $x$, $\psi(x, x) = 0$, $\psi(x, y)$ is strictly increasing in $x$ and strictly decreasing in $y$.

If in some event $X$ yields $x$ and $Y$ yields $y$, then $\psi(x, y)$ is a measure of the decision maker’s ex post feelings about the choice of $X$ over $Y$. This leads to the next definition.

\noindent\textbf{Definition 2.} Let $X, Y \in L$, where $X = (x_1, S_1; \ldots; x_n, S_n)$ and $Y = (y_1, S_1; \ldots; y_n, S_n)$. The \textit{regret lottery} evaluating the choice of $X$ over $Y$ is
\[
\Psi(X, Y) = (\psi(x_1, y_1), p_1; \ldots; \psi(x_n, y_n), p_n),
\]
where $p_i = P(S_i)$, $i = 1, \ldots, n$. 

Refer to $\psi$ and $\Psi$ as a regret function and a regret lottery. The main model follows.

\noindent\textbf{Definition 3.} The preference relation $\succsim$ is \textit{regret based} if there exists a regret function $\psi$ and a continuous functional $V$ that is defined over regret lotteries such that for any $X, Y \in L$,
\[
X \succsim Y \text{ if and only if } V(\Psi(X, Y)) \geq 0.
\]

\section{Counterexamples}
\begin{proposition}[Bikhchandani \& Segal 2011] Let $\succsim$ be transitive, continuous, and regret based. For all $X$ and $Y$, if they have the same cumulative distribution with respect to $P$, then $X\sim Y$. 
\end{proposition}

One difference between Theorem 1 and Proposition 1 in Bikhchandani\& Segal (2011) is that Proposition 1 \textit{drops} two axioms: monotonicity and completeness. The following counterexamples imply that the assumptions for Proposition 1 is too weak due to the lack of completeness axiom. 

Assume $\psi(x,y)=x-y$, a continuous regret function.

\noindent\textit{Counterexample 1:
} Under the constant representation $V(\Psi(X,Y))=-1$, $X\not\succsim Y$ if $X,Y$ have the same distribution.

It will be proven that $V$ satisfies transitivity, continuity, and regret-based.

\begin{proof} First, observe $V(\Psi(X,Y))=-1$ for any regret lotteries. $V$ is regret-based because a constant function is continuous and $\psi(x,y)=x-y$ is a continuous regret function. 

Transitivity will be proven by contradiction. Assume transitivity does not hold; then, there must exist $X,Y,Z$ such that $X\succsim Y$, $Y\succsim Z$, and $X\not\succsim Z$. However, $V(\Psi(X,Y))<0$ implies $X\not\succsim Y$, a contradiction! So transitivity holds.

Since for all $X,Y:$ $V\not\geq 0$, we have $X\not\succsim Y$ for any $X,Y$. The upper contour sets and lower contour sets of $\succsim$ are all empty sets and thus closed sets, satisfying continuity.
\end{proof}

\noindent\textit{Counterexample 2:} Under the representation $V(\Psi(X,Y))=-\sum_i|\psi(x_i,y_i)p_i|$, $X\not\succsim Y$ if $X,Y$ have the same distribution and $X\neq Y$.

\begin{proof} First, observe that $-\sum_i|\psi(x_i,y_i)p_i|$ is regret-based because the sum of absolute value functions is continuous and $\psi(x,y)=x-y$ is a continuous regret function. 

Assume $X\succsim Y$ and $Y\succsim Z$. Observe that $-\sum_i|\psi(x_i,y_i)p_i|\leq 0$ and the equality holds if $\psi(x_i,y_i)=0$ for all $i$. Since $X\succsim Y$ implies $V(\Psi(X,Y))\geq 0$, we have $V(\Psi(X,Y))=0$. Thus, $x_i=y_i$ for all $i$ and $X=Y$. Similarly, $Y=Z$. Therefore, $X\succsim Z$. Transitivity holds.

Observe that the upper contour set $\{X:X\succsim Y\}=\{X\}$ is a singleton and that the lower contour set is also a singleton. Therefore, the upper contour sets and the lower contour sets of $\succsim$ are singleton sets and therefore closed sets, satisfying continuity.
\end{proof}

 Proposition 1 is the first step in Bikhchandani\& Segal (2011)'s proof (page 98) for Theorem 1, and is applied in their Section 3.3, page 100.

\nocite{*}
\printbibliography

\end{document}